\documentclass[preprint,showpacs,preprintnumbers,amsmath,amssymb]{revtex4}

\usepackage{graphicx}
\usepackage{dcolumn}
\usepackage{bm}

\begin{document}

\preprint{APS/123-QED}

\title{Temperature dependence of the nonlocal voltage in an Fe/GaAs electrical spin injection device}
\author{G. Salis}
\email{gsa@zurich.ibm.com}
\author {A. Fuhrer}
\author {R. R. Schlittler}
\author {L. Gross}
\author {S. F. Alvarado}

\affiliation{IBM Research - Zurich,
S\"aumerstrasse 4, 8803 R\"uschlikon, Switzerland}

\date{April 7, 2010}

\begin{abstract}
The nonlocal spin resistance is measured as a function of temperature in a Fe/GaAs spin-injection device. For nonannealed samples that show minority-spin injection, the spin resistance is observed up to room temperature and decays exponentially with temperature at a rate of 0.018\,K$^{-1}$. Post-growth annealing at 440\,K increases the spin signal at low temperatures, but the decay rate also increases to 0.030\,K$^{-1}$. From measurements of the diffusion constant and the spin lifetime in the GaAs channel, we conclude that sample annealing modifies the temperature dependence of the spin transfer efficiency at injection and detection contacts. Surprisingly, the spin transfer efficiency increases in samples that exhibit minority-spin injection.
\end{abstract}

\maketitle
The efficient injection of spin-polarized electrons from a ferromagnetic source into a semiconducting channel is a fundamental ingredient of spin-based electronic device concepts. The injected spin polarization can be detected by analyzing the degree of circular polarization of photons that are emitted in the semiconductor after recombination of the injected electrons with resident holes~\cite{Alvarado1992,Ohno1999,Fiederling1999,Zhu2001,Hanbicki2002}. In all-electrical devices, the concept of nonlocal spin detection~\cite{Johnson1985,Jedema2001,Lou2007,Ciorga2009,Erve2007,Appelbaum2007,Tombros2007} has been used to convert the injected spin polarization into a nonlocal voltage $\Delta U_{\rm nl}$ that is measured at a ferromagnetic detection contact to which the electron spins diffuse. This voltage depends not only on how efficient spins are injected and detected, but also on the loss of spin polarization during the diffusive spin transport in the semiconductor. For electrical spin-injection into GaAs, a rapid decay of $\Delta U_{\rm nl}$ with temperature has been reported~\cite{Lou2007,Ciorga2009}, in contrast to the measured circular polarization of electroluminescence that remains observable up to room temperature~\cite{Zhu2001} and is strongly influenced by the interplay of spin lifetime and radiative recombination time~\cite{Salis2005b}. For Fe on GaAs(001), the sign and magnitude of the measured spin injection depend delicately on the growth temperature of the Fe layer as well as on a post-growth annealing treatment, where a reversal from minority to majority spin injection has been observed~\cite{Schultz2009}. Post-growth annealing also has a strong influence on the magnetic properties of ferromagnetic thin films on III-V compounds~\cite{Shaw2007,Bianco2008}.

Here we report a considerable change in the temperature dependence of the spin transfer efficiency across the ferromagnet/semiconductor interface that occurs after annealing Fe/GaAs samples at moderate temperature. We investigate the nonlocal spin resistance $\Delta \rho_{\rm nl}=\partial \Delta U_{\rm nl}/\partial I$ as a function of temperature $T$ up to 300\,K ($I$ is the current across the spin injection contact). We find an exponential decay of $\Delta \rho_{\rm nl}$ with $T$ from 5 to 200\,K with a rate that depends strongly on the annealing conditions. After annealing the sample at 440\,K, $\Delta \rho_{\rm nl}$ decays considerably faster with $T$. In order to understand this $T$ dependence, we characterize the spin decay $S(T)$ in the GaAs channel by measuring the spin lifetime $\tau_s$ and the diffusion constant $D$ in the channel. Since $S(T)$ does not change after sample annealing, the strong modification with annealing must arise from a change in the $T$-dependence of $\eta_i \eta_d$. By writing $\Delta \rho_{\rm nl}\propto S\eta_i\eta_d$, where the spin-injection efficiency $\eta_i$ specifies the spin polarization of an electron that has just tunneled from the Fe injection contact into the GaAs channel, and $\eta_d$ describes the relation between the spin polarization below the detection contact and $\Delta U_{\rm nl}$, we find the $T$ dependence of $\eta_i\eta_d$. For annealed samples, $\eta_i\eta_d$ is almost independent on $T$, whereas for those nonannealed samples that show minority-spin injection, $\eta_i \eta_d$ rises with $T$ between 30 and 140\,K. This unexpected result is discussed in terms of interface-related mechanisms that increasingly favor minority-spin transfer across the interface.

\begin{figure}[ht]
\includegraphics[width=80mm]{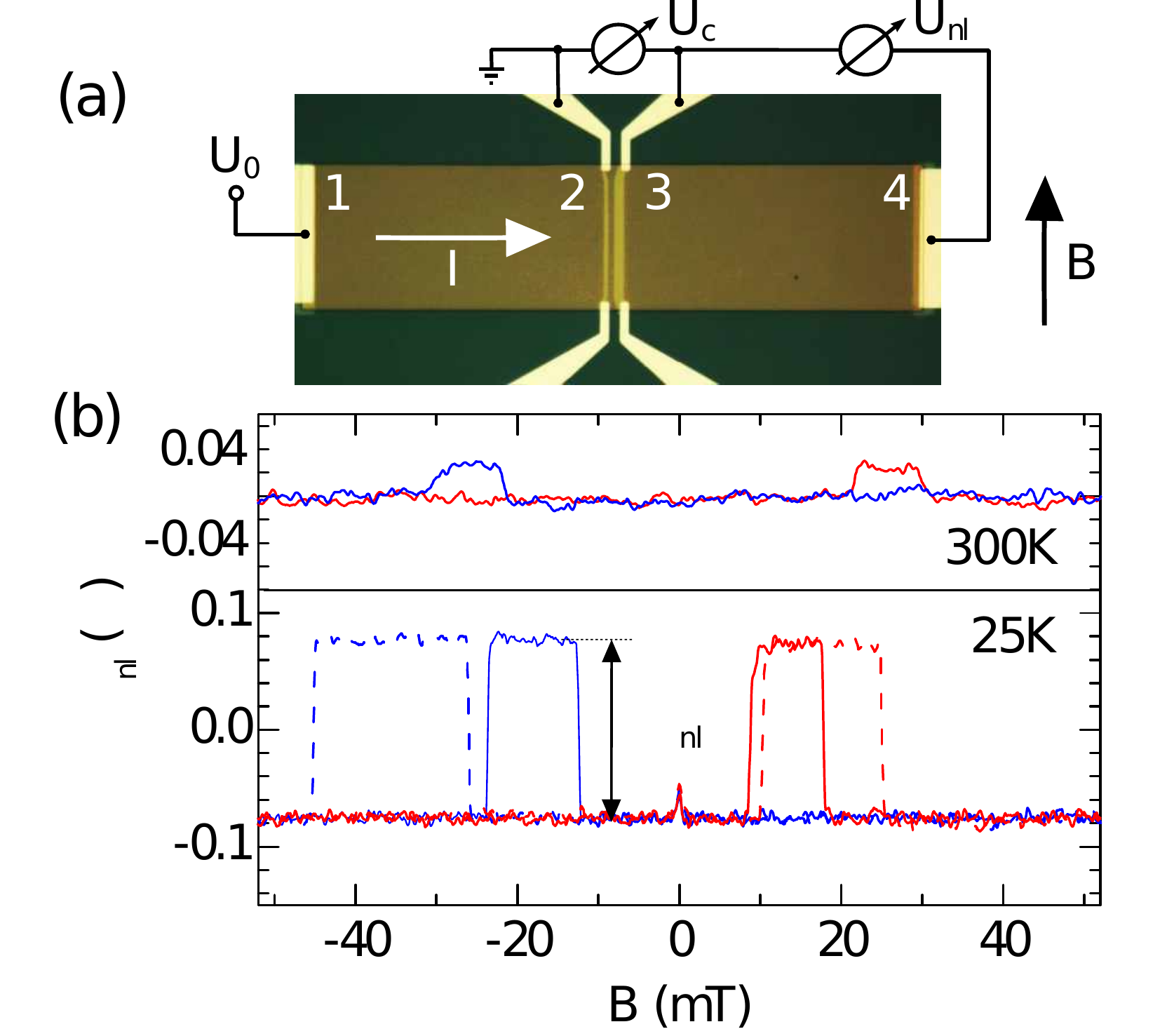}
\caption{\label{fig:fig1} (Color online) (a) Microscope image of the device consisting of four Fe bars, with the two inner bars serving as injection and detection  contacts, whereas the outer bars are used as reference contacts for spin injection and for measurement of $U_{\rm nl}$. (b) Measured nonlocal resistance $\rho_{\rm nl}$ for a nonannealed sample at $U_{\rm c}=-100$\,mV. At 25\,K, the magnetization reversal (steps $\Delta \rho_{\rm nl}$) occurs at stochastic switching fields, and single up (red) and down (blue) sweeps are shown. At higher $T$, repeatable switching fields are observed. Data at 300\,K are averaged over 10 sweeps. A background was subtracted that is linear in $B$.}
\end{figure}

The samples under investigation consist of an n-doped GaAs spin transport layer as described in Ref.~\cite{Salis2009} and Fe contacts for spin injection and detection. Four 56-$\mu$m-long contact bars were defined by evaporating 5\,nm Fe and 2\,nm Au through a nanostencil mask~\cite{Zahl2005}, see Fig.~\ref{fig:fig1}(a). These bars are electrically contacted by 100-nm-thick TiAu that is insulated from GaAs by a 100-nm-thick Al$_2$O$_3$ layer. The middle bars (2 and 3) serving as spin injection and detection contacts are 1 and 3 $\mu$m wide and separated by a gap of $a=2.4$\,$\mu$m. Between contacts 1 and 2, a bias $U_0$ is applied. For positive $U_0$, spin-polarized electrons are injected at contact 2 into the spin transport channel and drift towards contact 1. For negative $U_0$, electrons drift from contact 1 to contact 2 and spin polarization accumulates below contact 2 because of spin filtering~\cite{Lou2007}. In both cases, spin polarization diffuses towards contact 3, where a potential $U_{\rm nl}$ with respect to contact 4 is measured. By switching the alignment of the magnetizations of the injection and detection contacts from antiparallel to parallel with an external magnetic field $B$, two values for $U_{\rm nl}$ are measured that differ by $\Delta U_{\rm nl}$ which is proportional to the average spin polarization below the detection contact. From measurements using a lock-in amplifier, we obtain $\rho_{\rm nl}=\partial U_{\rm nl}/\partial I$, from which $U_{\rm nl}$ and $\Delta U_{\rm nl}$ were determined by integration over $I$. The voltage drop $U_{\rm c}$ across the Schottky barrier of contact 2 was measured as a potential difference between contacts 3 and 2. The highest temperature that the nonannealed samples have seen before measurement is 390\,K.

\begin{figure}[ht]
\includegraphics[width=80mm]{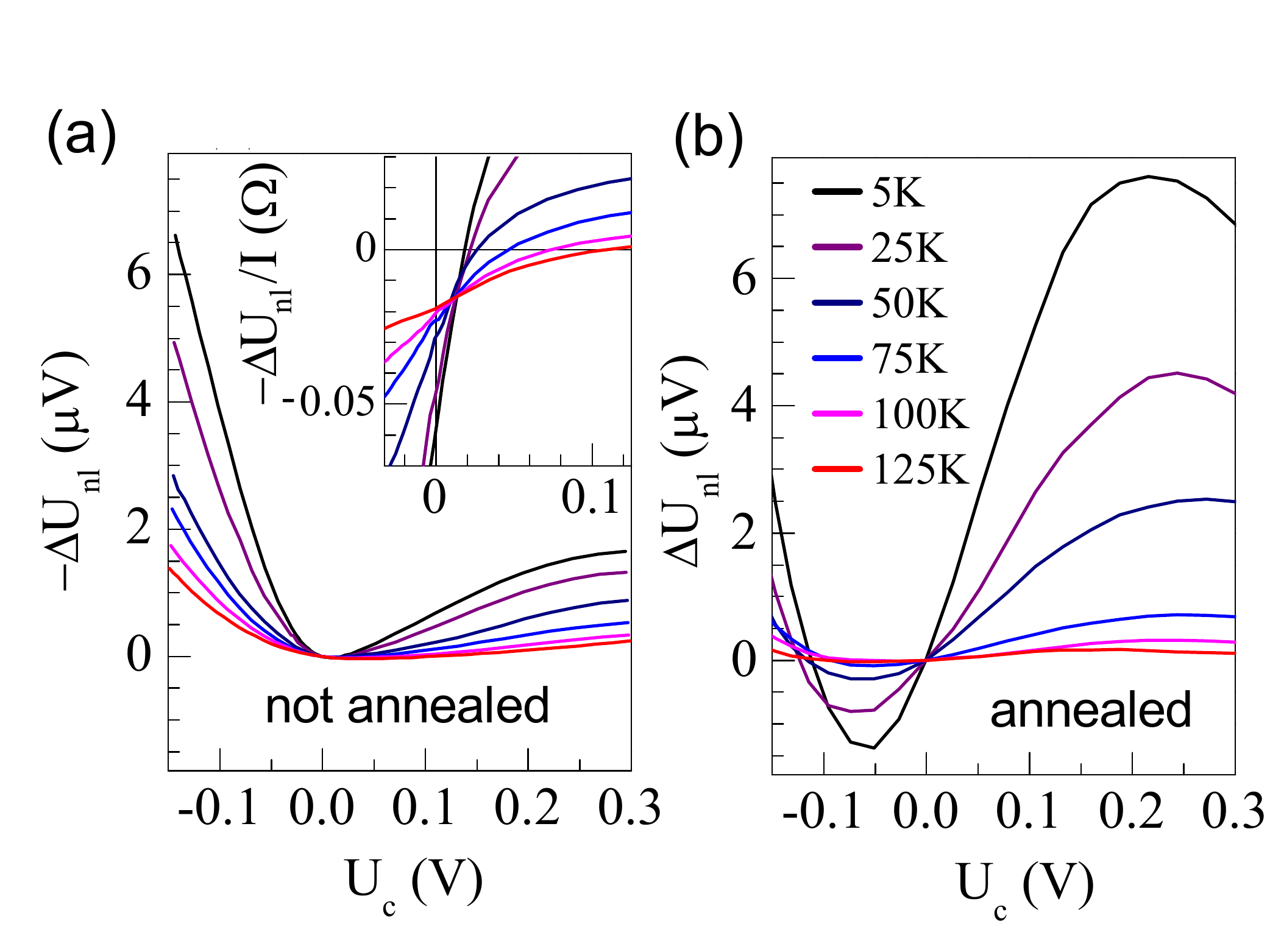}
\caption{\label{fig:fig2} (Color online) Bias dependence of $\Delta U_{\rm nl}$ vs. $U_{\rm c}$, for different $T$ before (a) and after (b) annealing. In nonannealed samples, a sign reversal of $\Delta U_{\rm nl}$ for positive $U_{\rm c}$ (spin injection) is observed whose position shifts to higher $U_{\rm c}$ as $T$ is increased, see inset of (a), which shows $-\Delta U_{\rm nl}/I$. Annealed samples exhibit a sign reversal of $U_{\rm nl}$ for negative $U_{\rm c}$.}
\end{figure}

From data of $\rho_{\rm nl}$ versus $B$ [Fig.~\ref{fig:fig1}(b)] we determine $\Delta \rho_{\rm nl}$. For nonannealed samples, $\Delta \rho_{\rm nl}$ decreases by a factor of 80 between 5 and 300\,K. The values for $\Delta U_{\rm nl}$ versus $U_{\rm c}$ are shown in Fig.~\ref{fig:fig2} for $T$ between 5 and 125\,K. We find a nonmonotonic dependence of $\Delta U_{\rm nl}$ on $U_{\rm c}$. It is helpful to consider that $\Delta U_{\rm nl}$ is proportional to the product of $I$ and the spin injection efficiency $\eta_i(U_{\rm c})$, such that $\eta_i \propto \Delta U_{\rm nl}/I$. As can be seen from the inset of Fig.~\ref{fig:fig2}(a), in the nonannealed sample, $\eta_i$ changes sign for $U_{\rm c}>0$, i.e., for spin injection. Such behavior has been related to a transition from minority to majority spin injection with increasing $U_{\rm c}$~\cite{Moser2006,Lou2007}. In the annealed samples, $\eta_i$ reverses its sign at $U_{\rm c}<0$, see Fig.~\ref{fig:fig2}(b), and majority spins are injected at $U_{\rm c}>0$. Because of the opposite sign of spin injection close to zero bias, also $\eta_d$ has opposite signs~\cite{Lou2007} for the the annealed and nonannealed sample. In Fig.~\ref{fig:fig2}, we therefore plot $-\Delta U_{\rm nl}$ for the nonannealed sample. Although the dependence of $\eta_i$ on $U_{\rm c}$ is not understood in detail, it has been related to the interfacial structure between Fe and GaAs~\cite{Chantis2007,Schultz2009} or to a confinement layer in the semiconductor~\cite{Dery2007}, which will be discussed later.

\begin{figure}[ht]
\includegraphics[width=80mm]{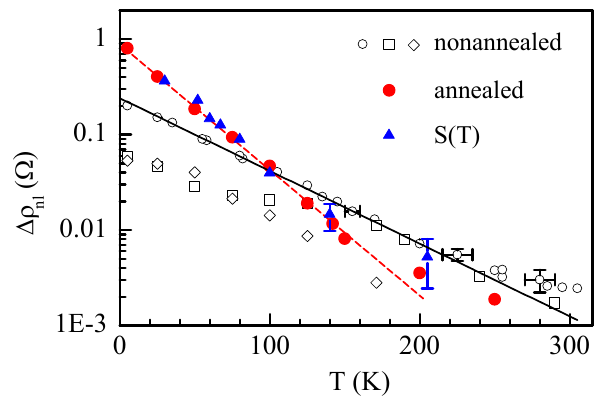}
\caption{\label{fig:fig3} (Color online) Temperature dependence of $\Delta \rho_{\rm nl}$ of the samples before and after annealing. For the nonannealed samples, data for $U_{\rm c}=-100$ (circles), 0 (squares) and 200\,mV (diamonds) are given. The solid and dashed lines are exponential decays with rates of 0.018 and 0.030\,K$^{-1}$ for the nonannealed and annealed samples, respectively. Filled (blue) triangles represent the expected signal as calculated using $D$ from Hall measurements and $\tau_s$ from one-parameter Hanle fits.}
\end{figure}

Figure~\ref{fig:fig3} summarizes the $T$-dependence of $\Delta \rho_{\rm nl}$ for both nonannealed and annealed samples. For the nonannealed sample, $\Delta \rho_{\rm nl}$ decays exponentially with increasing $T$ up to 200\,K. The decay rate is about $0.018$\,K$^{-1}$ (black line) for both $U_{\rm c}=-200$ and 100\,mV. For $U_{\rm c}=0$\,mV, a more complicated behavior with $T$ is observed, which we attribute to the crossing of $\Delta U_{\rm nl}=0$ at small positive $U_{\rm c}$, i.e. to the bias dependence of $\eta_i$. Annealing the sample increases $\Delta \rho_{\rm nl}$ at low $T$, but at the same time, $\Delta \rho_{\rm nl}$ decreases much faster with $T$, namely, at a rate of 0.030\,K$^{-1}$ (red dashed line). This faster decrease results in a signal that at higher $T$ becomes smaller than that of the nonannealed sample. The data for the annealed sample were obtained at $U_{\rm c}=0$, but similar behavior is observed for $U_{\rm c}=100$\,mV (not shown).

\begin{figure}[ht]
\includegraphics[width=80mm]{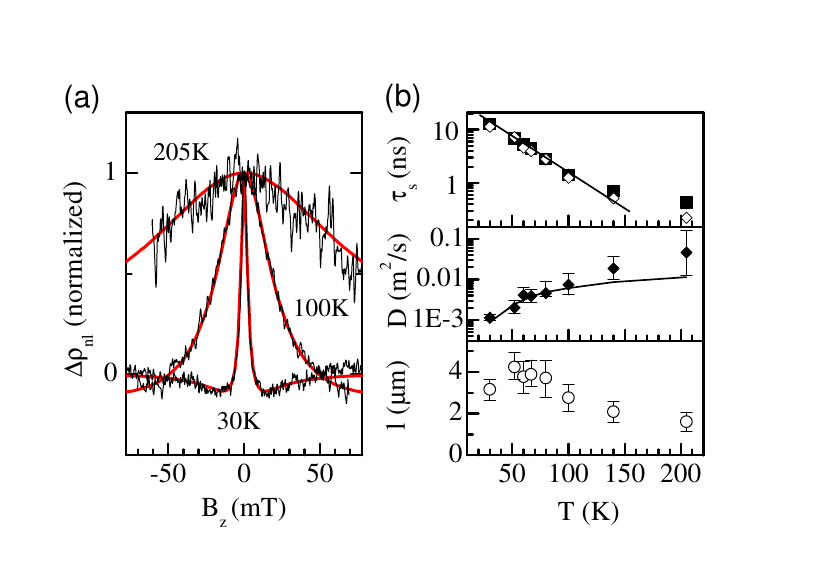}
\caption{\label{fig:fig4} (Color online) (a) Hanle measurements of $\Delta \rho_{\rm nl}$ at different temperatures (thin black line) and two-parameter fits (thick red line) on a nonannealed sample. A magnetic field $B_z$ perpendicular to the sample plane was swept at $U_{\rm c}=-100$\,mV, and plotted are the differences between sweeps with antiparallel and parallel magnetization of contacts 2 and 3. (b) Parameters obtained from the Hanle fits. The spin lifetime $\tau_s$ results from one and two-parameter Hanle fits (open diamonds and filled squares), the diffusion constant $D$ from Hanle fits (diamonds) and from Hall measurements (line). The error bars indicate the uncertainty from a background subtraction from the Hanle signal. The spin diffusion length $l$ is calculated from $D$ and $\tau_s$.}
\end{figure}

The magnitude of $\Delta \rho_{\rm nl}$ depends on $\eta_i$ and $\eta_d$, as well as on the transport and spin dynamics in the GaAs channel that reduces the injected spin polarization to a value $S$ at the detection contact. $S$ is characterized by the diffusion constant $D$ and the spin lifetime $\tau_s$, which are obtained from Hanle measurements at different $T$, as shown in Fig.~\ref{fig:fig4}(a). When the spins in the channel precess about a perpendicular magnetic field $B_z$, $\Delta \rho_{\rm nl}$ decreases because of the distribution in the arrival times of the injected spins at contact 3. This can be calculated by the one-dimensional Hanle integral~\cite{Jedema2001}

\begin{equation}
\label{eq1}
\Delta \rho_{\rm nl}\propto \int_0^{\infty}{1\over{\sqrt{4\pi D t}}}e^{-{x^2\over{4Dt}}}\cos \left(g\mu_B B_z t\over{\hbar}\right)e^{-t/\tau_s}dt,
\end{equation}

\noindent where $g=0.44$ is the electron g-factor of GaAs, $\mu_B$ the Bohr magneton, $\hbar$ Planck's constant, and $x$ the distance between injection and detection of the spins. To account for the finite width $w_3$ of the detection contact, Eq.~(\ref{eq1}) is integrated for $x$ ranging from $a$ to $a+w_3$. Because of the electric field applied between contacts 1 and 2, it is assumed that all spins are injected at the edge of contact 2 towards contact 3.

Two-parameter fits of the Hanle data with $\tau_s$ and $D$ as parameters are shown in Fig.~\ref{fig:fig4}(a), and the resulting fit parameters are summarized in Fig.~\ref{fig:fig4}(b). The error bar for $D$ arises from an uncertainty in an offset in $\Delta \rho_{\rm nl}(B_z)$, which becomes larger at higher $T$, where the tails of the Hanle peak can no longer be measured. For $T>70$\,K, the Hanle peak was normalized to the value of $\Delta \rho_{\rm nl}$ for an in-plane magnetic-field sweep. For $T<100$\,K, the values of $D$ of the two-parameter Hanle fits match well the data of $D$ obtained from a four-point Hall measurement. For $T>100$\,K, the Hanle fit values deviate towards higher values, which we attribute to the weak influence of $D$ on the Hanle curves at higher $T$. We therefore also fit the data with $\tau_s$ as the only fit parameter and fix $D$ to the transport value. In Fig.~\ref{fig:fig4}(b), the results for $\tau_s$ are shown for both the one- and two-parameter Hanle fits. The two fits yield similar results, namely, a $\tau_s$ that decays approximately exponentially with $T$ from 12 ns at 30\,K to 200\,ps at 205\,K. From $D$ and $\tau_s$, the spin diffusion length $l=\sqrt{D \tau_s}$ is calculated. The combined increase of $D$ and decrease of $\tau_s$ lead to only a small decrease of $l$, from 4\,$\mu$m at 50\,K to 1.6\,$\mu$m at 200\,K. In Ref.~\onlinecite{Ciorga2009}, $l=2.8$\,$\mu$m was found at $T=4$\,K, whereas Ref.~\onlinecite{Lou2007} measures $l=6$\,$\mu$m at 50\,K, comparable to our values. It is important to note that post-growth annealing does not affect the spin transport properties of the GaAs channel, as verified in separate measurements on an annealed sample.

The contribution $S$ to $\Delta \rho_{\rm nl}$ can be calculated from Eq.~(\ref{eq1}) with $B_z=0$. The integration over $t$ yields $S \propto (\tau_s/l)\exp({-x/l})$. $S$ decays exponentially with the separation $x$ between injection and the detection contact, and $\tau_s/l$ stems from the integration over time, where spins contribute in a time $\tau_s$ and spread over a length $l$. The proportionality of $S$ to $\tau_s$ influences $S(T)$ more strongly than the relatively weak variation of $l$ with $T$ does. $S(T)$ as obtained~\cite{note2010_2} from the one-parameter Hanle fits is shown in Fig.~\ref{fig:fig3} as triangles, scaled by a factor for better comparison with $\Delta \rho_{\rm nl}$. The overall $T$-dependence of $\Delta \rho_{\rm nl}$ is determined from $\eta_i \eta_d S(T)$. Since the decay rate of $\Delta \rho_{\rm nl}$ of the annealed sample is very similar to that of $S(T)$, $\eta_i \eta_d$ does not change much with $T$ in that sample. However, before annealing, $\Delta \rho_{\rm nl}$ decreases significantly less, amounting to a factor of 3.7 between 30 and 140\,K. This suggests that $\eta_i \eta_d$ increases with $T$ below 140\,K. Above 140\,K, the slope of $\Delta \rho_{\rm nl}(T)$ is similar to that of $S(T)$ for both the annealed and the nonannealed samples.

To ensure that Eq.~(\ref{eq1}) and thus $S(T)$ does not overestimate the decay rate attributed to the GaAs channel, some care has to be taken. In fact, there are several limitations to Eq.~(\ref{eq1}). First, it is derived in the limit of small spin polarization in the GaAs channel by assuming that there the spin injection rate does not depend on the spin polarization in the GaAs channel. More generally, the spin injection rate is proportional to the difference between $\eta_i$ and the spin polarization in the GaAs channel. This modification is equivalent to adding an effective spin decay rate 1/$\tau_0$, given by the rate of injected electrons divided by the number of electrons in the channel below the injection contact. For our sample geometry, we obtain $\tau_0 \approx 10$\,ns for a typical current of $I=50$\,$\mu$A, which is comparable to the spin lifetime of 12\,ns at 30\,K. The result is that $S$ saturates with increasing $\tau_s$. From a solution of the spin drift-diffusion equation, we find that the corresponding reduction of $S$ is smaller than 30\% at 30\,K. Second, Eq.~(\ref{eq1}) neglects the drift of spin polarization towards contact 1. This effect has an influence on $S$ that is smaller than 10\% for $I=50$\,$\mu$A at $U_{\rm c}=-100$\,mV and at 30\,K. We can also neglect that dynamic nuclear polarization enhances the applied field~\cite{Lou2006,Salis2009} at small $T$, which would lead to overestimated values for $\tau_s$. Such dynamic nuclear polarization sensitively depends on a misalignment between sample normal and $B_z$, and we do not find large variations of the measured $\tau_s$ above 30\,K.

Considering the limitations above, we estimate that the decay rate of $S$ between 30 and 140\,K is at least 0.025\,K$^{-1}$, corresponding to a loss of spin polarization in the channel of a factor of at least 16 when $T$ is increased from 30 to 140\,K. In this $T$-range, however, $\Delta \rho_{\rm nl}$ measured in the nonannealed sample exhibits a much weaker decrease, namely, only by a factor of 7, whereas for the annealed sample the factor increases to 27. These strong, annealing-induced changes must be related to a modification of the spin injection or detection efficiency that occurs at the interface. They are concomitant with a shift from minority- to majority-spin injection. We recall that annealing at moderate $T$ is known to affect only the Fe/GaAs(001) interface region, giving rise to structural changes that lead to an enhanced crystal order~\cite{Zega2006,Shaw2007} and enhanced polarization injection efficiency~\cite{Zega2006,Adelmann2005}. The important result is that before moderate post-growth annealing, $\eta_i \eta_d$ increases with increasing $T$. Such an increase is likely to be related to a change in the weighting of minority- and majority-spin processes~\cite{note2010_1}. The spin-filter effect due to symmetry conservation of the coupling between the Fe and GaAs wave functions at the interface strongly favors majority spin injection~\cite{Wunnicke2002} for the case of well-ordered Fe/GaAs interfaces. On the other hand, a resonant state arising from interface layers promotes minority-spin injection~\cite{Chantis2007}. In addition, the existence of a bound state in the semiconductor close to the interface can influence both the size and sign of accumulated spin polarization~\cite{Dery2007}. The exact balance of minority and majority spin contributions is determined by the interplay of all these effects. The important question is how a change in $T$ redistributes the weight of the individual contributions. As $T$ is increased, the distribution of electrons that tunnel through the Schottky barrier extends towards higher energies at which the tunneling probability becomes significantly larger. This change in energy and a simultaneous modification of the in-plane electron momentum of the tunneling electrons may influence the resonance with the minority peak~\cite{Chantis2007}. An additional role could be played by a change in the overlap of GaAs and Fe wave-functions that determines the spin filtering efficiency, as well as by a $T$-dependent variation of the occupation of a semiconductor bound state. The similar decay rate found for positive and negative $U_c$ in the nonannealed sample points to a positive $T$-dependence of $\eta_d$, induced by a shift towards the minority peak as $T$ is increased. This is strongly supported by the position of the zero-crossing of $\Delta U_{\rm nl}$ that occurs at higher $U_c$ when $T$ is increased [inset of Fig.~\ref{fig:fig2}(a)], which is compatible with the majority contribution moving farther away from $U_c=0$ and therefore the minority peak gaining in strength. This results in a positive $T$-dependence of $\eta_i$ at $U_c=0$ and by reciprocity also of $\eta_d$. Away from $U_c$, only the increase in $\eta_d$ is seen in $\Delta \rho_{\rm nl}$, whereas at $U_c=0$, both $\eta_i$ and $\eta_d$ contribute, leading to an even slower decay of $\Delta \rho_{\rm nl}$, in agreement with our data, see Fig.~\ref{fig:fig3}.

In conclusion, we observe a strong change in the temperature dependence of the nonlocal spin resistance $\Delta \rho_{\rm nl}$ upon post-growth sample annealing, which we relate to a change in the $T$ dependence of the spin-transfer efficiency at the injection and detection contacts. For annealed samples, the spin transfer efficiencies increase with $T$ up to 140\,K. Such behavior is likely associated with an interplay of interface-related mechanisms that more and more favors the minority-spin as $T$ is increased.

We acknowledge valuable discussions with Rolf Allenspach, and technical support from Meinrad Tschudy, Daniele Caimi, Ute Drechsler and Martin Witzig.


\begin{thebibliography}{27}
\expandafter\ifx\csname natexlab\endcsname\relax\def\natexlab#1{#1}\fi
\expandafter\ifx\csname bibnamefont\endcsname\relax
  \def\bibnamefont#1{#1}\fi
\expandafter\ifx\csname bibfnamefont\endcsname\relax
  \def\bibfnamefont#1{#1}\fi
\expandafter\ifx\csname citenamefont\endcsname\relax
  \def\citenamefont#1{#1}\fi
\expandafter\ifx\csname url\endcsname\relax
  \def\url#1{\texttt{#1}}\fi
\expandafter\ifx\csname urlprefix\endcsname\relax\def\urlprefix{URL }\fi
\providecommand{\bibinfo}[2]{#2}
\providecommand{\eprint}[2][]{\url{#2}}

\bibitem[{\citenamefont{Alvarado and Renaud}(1992)}]{Alvarado1992}
\bibinfo{author}{\bibfnamefont{S.~F.} \bibnamefont{Alvarado}} \bibnamefont{and}
  \bibinfo{author}{\bibfnamefont{P.}~\bibnamefont{Renaud}},
  \bibinfo{journal}{Phys. Rev. Lett.} \textbf{\bibinfo{volume}{68}},
  \bibinfo{pages}{1387} (\bibinfo{year}{1992}).

\bibitem[{\citenamefont{Ohno et~al.}(1999)\citenamefont{Ohno, Young, Beschoten,
  Matsukura, Ohno, and Awschalom}}]{Ohno1999}
\bibinfo{author}{\bibfnamefont{Y.}~\bibnamefont{Ohno}}
  \bibinfo{author}{\bibfnamefont{et~al.}},
  \bibinfo{journal}{Nature} \textbf{\bibinfo{volume}{402}},
  \bibinfo{pages}{790} (\bibinfo{year}{1999}).

\bibitem[{\citenamefont{Fiederling et~al.}(1999)\citenamefont{Fiederling, Keim,
  Reuscher, Ossau, Schmidt, Waag, and Molenkamp}}]{Fiederling1999}
\bibinfo{author}{\bibfnamefont{R.}~\bibnamefont{Fiederling}}
  \bibinfo{author}{\bibfnamefont{et~al.}},
  \bibinfo{journal}{Nature} \textbf{\bibinfo{volume}{402}},
  \bibinfo{pages}{787} (\bibinfo{year}{1999}).

\bibitem[{\citenamefont{Zhu et~al.}(2001)\citenamefont{Zhu, Ramsteiner,
  Kostial, Wassermeier, Sch\"{o}nherr, and Ploog}}]{Zhu2001}
\bibinfo{author}{\bibfnamefont{H.~J.} \bibnamefont{Zhu}}
  \bibinfo{author}{\bibfnamefont{et~al.}},
  \bibinfo{journal}{Phys. Rev. Lett.} \textbf{\bibinfo{volume}{87}},
  \bibinfo{pages}{016601} (\bibinfo{year}{2001}).

\bibitem[{\citenamefont{Hanbicki et~al.}(2002)\citenamefont{Hanbicki, Jonker,
  Itskos, Kioseoglou, and Petrou}}]{Hanbicki2002}
\bibinfo{author}{\bibfnamefont{A.~T.} \bibnamefont{Hanbicki}}
  \bibinfo{author}{\bibfnamefont{et~al.}},
  \bibinfo{journal}{Appl. Phys. Lett.} \textbf{\bibinfo{volume}{80}},
  \bibinfo{pages}{1240} (\bibinfo{year}{2002}).

\bibitem[{\citenamefont{Johnson and Silsbee}(1985)}]{Johnson1985}
\bibinfo{author}{\bibfnamefont{M.}~\bibnamefont{Johnson}} \bibnamefont{and}
  \bibinfo{author}{\bibfnamefont{R.~H.} \bibnamefont{Silsbee}},
  \bibinfo{journal}{Phys. Rev. Lett.} \textbf{\bibinfo{volume}{55}},
  \bibinfo{pages}{1790} (\bibinfo{year}{1985}).

\bibitem[{\citenamefont{Jedema et~al.}(2001)\citenamefont{Jedema, Filip, and
  van Wees}}]{Jedema2001}
\bibinfo{author}{\bibfnamefont{F.~J.} \bibnamefont{Jedema}},
  \bibinfo{author}{\bibfnamefont{A.~T.} \bibnamefont{Filip}}, \bibnamefont{and}
  \bibinfo{author}{\bibfnamefont{B.~J.} \bibnamefont{van Wees}},
  \bibinfo{journal}{Nature} \textbf{\bibinfo{volume}{410}},
  \bibinfo{pages}{345} (\bibinfo{year}{2001}).

\bibitem[{\citenamefont{Lou et~al.}(2007)\citenamefont{Lou, Adelmann, Crooker,
  Garlid, Zhang, Reddy, Flexner, Palmstrom, and Crowell}}]{Lou2007}
\bibinfo{author}{\bibfnamefont{X.}~\bibnamefont{Lou}}
  \bibinfo{author}{\bibfnamefont{et~al.}},
  \bibinfo{journal}{Nat. Phys.}
  \textbf{\bibinfo{volume}{3}}, \bibinfo{pages}{197} (\bibinfo{year}{2007}).

\bibitem[{\citenamefont{Ciorga et~al.}(2009)\citenamefont{Ciorga, Einwanger,
  Wurstbauer, Schuh, Wegscheider, and Weiss}}]{Ciorga2009}
\bibinfo{author}{\bibfnamefont{M.}~\bibnamefont{Ciorga}}
  \bibinfo{author}{\bibfnamefont{et~al.}},
  \bibinfo{journal}{Phys. Rev. B}
  \textbf{\bibinfo{volume}{79}}, \bibinfo{eid}{165321}
  (\bibinfo{year}{2009}).

\bibitem[{\citenamefont{van~'t Erve et~al.}(2007)\citenamefont{van~'t Erve,
  Hanbicki, Holub, Li, Awo-Affouda, Thompson, and Jonker}}]{Erve2007}
\bibinfo{author}{\bibfnamefont{O.~M.~J.} \bibnamefont{van~'t Erve}}
  \bibinfo{author}{\bibfnamefont{et~al.}},
  \bibinfo{journal}{Appl. Phys. Lett.}
  \textbf{\bibinfo{volume}{91}}, \bibinfo{eid}{212109} (\bibinfo{year}{2007}).

\bibitem[{\citenamefont{Appelbaum et~al.}(2007)\citenamefont{Appelbaum, Huang,
  and Monsma}}]{Appelbaum2007}
\bibinfo{author}{\bibfnamefont{I.}~\bibnamefont{Appelbaum}},
  \bibinfo{author}{\bibfnamefont{B.}~\bibnamefont{Huang}}, \bibnamefont{and}
  \bibinfo{author}{\bibfnamefont{D.~J.} \bibnamefont{Monsma}},
  \bibinfo{journal}{Nature} \textbf{\bibinfo{volume}{447}},
  \bibinfo{pages}{295} (\bibinfo{year}{2007}).

\bibitem[{\citenamefont{Tombros et~al.}(2007)\citenamefont{Tombros, Jozsa,
  Popinciuc, Jonkman, and van Wees}}]{Tombros2007}
\bibinfo{author}{\bibfnamefont{N.}~\bibnamefont{Tombros}}
  \bibinfo{author}{\bibfnamefont{et~al.}},
  \bibinfo{journal}{Nature} \textbf{\bibinfo{volume}{448}},
  \bibinfo{pages}{571} (\bibinfo{year}{2007}).

\bibitem[{\citenamefont{Salis et~al.}(2005)\citenamefont{Salis, Wang, Jiang,
  Shelby, Parkin, Bank, and Harris}}]{Salis2005b}
\bibinfo{author}{\bibfnamefont{G.}~\bibnamefont{Salis}}
  \bibinfo{author}{\bibfnamefont{et~al.}},
  \bibinfo{journal}{Appl. Phys. Lett.} \textbf{\bibinfo{volume}{87}},
  \bibinfo{eid}{262503} (\bibinfo{year}{2005}).

\bibitem[{\citenamefont{Schultz et~al.}(2009)\citenamefont{Schultz, Marom,
  Naveh, Lou, Adelmann, Strand, Crowell, Kronik, and
  Palmstr\o{}m}}]{Schultz2009}
\bibinfo{author}{\bibfnamefont{B.~D.} \bibnamefont{Schultz}}
  \bibinfo{author}{\bibfnamefont{et~al.}},
  \bibinfo{journal}{Phys. Rev. B} \textbf{\bibinfo{volume}{80}},
  \bibinfo{pages}{201309(R)} (\bibinfo{year}{2009}).

\bibitem[{\citenamefont{Shaw and Falco}(2007)}]{Shaw2007}
\bibinfo{author}{\bibfnamefont{J.~M.} \bibnamefont{Shaw}} \bibnamefont{and}
  \bibinfo{author}{\bibfnamefont{C.~M.} \bibnamefont{Falco}},
  \bibinfo{journal}{J. Appl. Phys.} \textbf{\bibinfo{volume}{101}},
  \bibinfo{eid}{033905} (\bibinfo{year}{2007}).

\bibitem[{\citenamefont{Bianco et~al.}(2008)\citenamefont{Bianco, Bouchon,
  Sousa, Salis, and Alvarado}}]{Bianco2008}
\bibinfo{author}{\bibfnamefont{F.}~\bibnamefont{Bianco}}
  \bibinfo{author}{\bibfnamefont{et~al.}},
  \bibinfo{journal}{J. Appl. Phys.} \textbf{\bibinfo{volume}{104}}
  (\bibinfo{year}{2008}).

\bibitem[{\citenamefont{Salis et~al.}(2009)\citenamefont{Salis, Fuhrer, and
  Alvarado}}]{Salis2009}
\bibinfo{author}{\bibfnamefont{G.}~\bibnamefont{Salis}},
  \bibinfo{author}{\bibfnamefont{A.}~\bibnamefont{Fuhrer}}, \bibnamefont{and}
  \bibinfo{author}{\bibfnamefont{S.~F.} \bibnamefont{Alvarado}},
  \bibinfo{journal}{Phys. Rev. B} \textbf{\bibinfo{volume}{80}},
  \bibinfo{pages}{115332} (\bibinfo{year}{2009}).

\bibitem[{\citenamefont{Zahl et~al.}(2005)\citenamefont{Zahl, Bammerlin, Meyer,
  and Schlittler}}]{Zahl2005}
\bibinfo{author}{\bibfnamefont{P.}~\bibnamefont{Zahl}}
  \bibinfo{author}{\bibfnamefont{et~al.}},
  \bibinfo{journal}{Rev. Sci. Instr.}
  \textbf{\bibinfo{volume}{76}}, \bibinfo{eid}{023707}
  (\bibinfo{year}{2005}).

\bibitem[{\citenamefont{Moser et~al.}(2006)\citenamefont{Moser, Zenger, Gerl,
  Schuh, Meier, Chen, Bayreuther, Wegscheider, Weiss, Lai et~al.}}]{Moser2006}
\bibinfo{author}{\bibfnamefont{J.}~\bibnamefont{Moser}}
  \bibinfo{author}{\bibfnamefont{et~al.}},
  \bibinfo{journal}{Appl. Phys. Lett.}
  \textbf{\bibinfo{volume}{89}}, \bibinfo{eid}{162106}
  (\bibinfo{year}{2006}).

\bibitem[{\citenamefont{Chantis et~al.}(2007)\citenamefont{Chantis,
  Belashchenko, Smith, Tsymbal, van Schilfgaarde, and Albers}}]{Chantis2007}
\bibinfo{author}{\bibfnamefont{A.~N.} \bibnamefont{Chantis}}
  \bibinfo{author}{\bibfnamefont{et~al.}},
  \bibinfo{journal}{Phys. Rev. Lett.}
  \textbf{\bibinfo{volume}{99}}, \bibinfo{pages}{196603}
  (\bibinfo{year}{2007}).

\bibitem[{\citenamefont{Dery and Sham}(2007)}]{Dery2007}
\bibinfo{author}{\bibfnamefont{H.}~\bibnamefont{Dery}} \bibnamefont{and}
  \bibinfo{author}{\bibfnamefont{L.~J.} \bibnamefont{Sham}},
  \bibinfo{journal}{Phys. Rev. Lett.} \textbf{\bibinfo{volume}{98}},
  \bibinfo{pages}{046602} (\bibinfo{year}{2007}).

\bibitem[{not({\natexlab{a}})}]{note2010_2}
\bibinfo{note}{For calculation of $S(T)$, an additional integration over the
  width of the detection contact did not change the slope of $S(T)$.}

\bibitem[{\citenamefont{Lou et~al.}(2006)\citenamefont{Lou, Adelmann, Furis,
  Crooker, Palmstr\o{}m, and Crowell}}]{Lou2006}
\bibinfo{author}{\bibfnamefont{X.}~\bibnamefont{Lou}}
  \bibinfo{author}{\bibfnamefont{et~al.}},
  \bibinfo{journal}{Phys. Rev. Lett.}
  \textbf{\bibinfo{volume}{96}}, \bibinfo{pages}{176603}
  (\bibinfo{year}{2006}).

\bibitem[{\citenamefont{Zega et~al.}(2006)\citenamefont{Zega, Hanbicki, Erwin,
  \ifmmode \check{Z}\else \v{Z}\fi{}uti\ifmmode~\acute{c}\else \'{c}\fi{},
  Kioseoglou, Li, Jonker, and Stroud}}]{Zega2006}
\bibinfo{author}{\bibfnamefont{T.~J.} \bibnamefont{Zega}}
  \bibinfo{author}{\bibfnamefont{et~al.}},
  \bibinfo{journal}{Phys. Rev. Lett.}
  \textbf{\bibinfo{volume}{96}}, \bibinfo{pages}{196101}
  (\bibinfo{year}{2006}).

\bibitem[{\citenamefont{Adelmann et~al.}(2005)\citenamefont{Adelmann, Lou,
  Strand, Palmstrøm, and Crowell}}]{Adelmann2005}
\bibinfo{author}{\bibfnamefont{C.}~\bibnamefont{Adelmann}}
  \bibinfo{author}{\bibfnamefont{et~al.}},
  \bibinfo{journal}{Phys. Rev. B}
  \textbf{\bibinfo{volume}{71}}, \bibinfo{pages}{121301(R)}
  (\bibinfo{year}{2005}).

\bibitem[{not({\natexlab{b}})}]{note2010_1}
\bibinfo{note}{Irrespective of sample annealing, $\eta_d$ also depends on the
  density-of-states in GaAs at the Fermi energy and thus on carrier density. In
  the measured $T$ range, we observe only a small increase of the carrier
  density, from 5 to 6\,$\times10^{16}$cm$^{-3}$.}

\bibitem[{\citenamefont{Wunnicke et~al.}(2002)\citenamefont{Wunnicke,
  Mavropoulos, Zeller, Dederichs, and Grundler}}]{Wunnicke2002}
\bibinfo{author}{\bibfnamefont{O.}~\bibnamefont{Wunnicke}}
  \bibinfo{author}{\bibfnamefont{et al.}},
  \bibinfo{journal}{Phys. Rev. B} \textbf{\bibinfo{volume}{65}},
  \bibinfo{pages}{241306(R)} (\bibinfo{year}{2002}).

\end{thebibliography}
\end{document}